\documentclass[aps,pra,superscriptaddress,a4paper,nofootinbib,floatfix,twocolumn,10pt]{revtex4-1}
\usepackage{graphicx,color,amsmath,amssymb,enumerate,verbatim}
\usepackage[normalem]{ulem}
\usepackage[ocgcolorlinks,colorlinks=true,linkcolor=blue,citecolor=red]{hyperref}

\newcommand{\tr}{\textrm{tr}}
\newcommand{\bra}[1]{\langle #1|}
\newcommand{\ket}[1]{|#1\rangle}

\newcommand{\kets}[2]{|#1\rangle^{_#2}}

\newcommand{\proj}[2]{|#1\rangle\!\langle#1|^{_#2}}

\newcommand{\identity}{\openone}

\newcommand{\assem}[3]{#1_{#2}^{#3}}

\newcommand{\rma}[0]{\mathrm{a}}

\newcommand{\rmA}[0]{\textsc{A}}
\newcommand{\bfa}[0]{\mathbf{a}}
\newcommand{\rmB}[0]{\textsc{B}}
\newcommand{\rmR}[0]{\mathrm{R}}

\newcommand{\nc}[0]{\mathrm{\o}}
\newcommand{\LHS}[0]{\textsc{lhs}}

\newcommand{\sectionprl}[1]{\section{#1}}

\newcommand{\ie}{{\it{i.e.}~}}

\begin{document}
\title{Loss-tolerant EPR steering for arbitrary dimensional states: joint measurability and unbounded violations under losses}
\author{Paul Skrzypczyk}\affiliation{H. H. Wills Physics Laboratory, University of Bristol$\text{,}$ Tyndall Avenue, Bristol, BS8 1TL, United Kingdom}
\affiliation{ICFO-Institut de Ciencies Fotoniques, Mediterranean Technology Park, 08860 Castelldefels, Barcelona, Spain}

\author{Daniel Cavalcanti}\affiliation{ICFO-Institut de Ciencies Fotoniques, Mediterranean Technology Park, 08860 Castelldefels, Barcelona, Spain}

\begin{abstract}
We show how to construct loss-tolerant linear steering inequalities using a generic set of von Neumann measurements that are violated by $d$-dimensional states, and that rely only upon a simple property of the set of measurements used (the maximal overlap between measurement directions). Using these inequalities we show that the critical detection efficiency above which $n$ von Neumann measurements can demonstrate steering is $1/n$. We show furthermore that using our construction and high dimensional states allows for steering demonstrations which are also highly robust to depolarising noise and produce unbounded violations in the presence of loss. Finally, our results provide an explicit means to certify the non-joint measurability of any set of inefficient von Neuman measurements.
\end{abstract}

\maketitle
Two fundamental aspects of quantum theory are entanglement and incompatibility of measurements, with both aspects lying at the heart of many applications in quantum information science. Interestingly, even in a scenario where neither the source of entanglement nor the measuring devices used are characterised, both the presence of entanglement and measurement incompatibility can be simultaneously certified. In the so called \emph{device-independent setting}, where no measuring device of any party is trusted, this is accomplished by the violation of a Bell inequality \cite{Bell64,Bell review}. In a \emph{semi-device-independent setting} where only a subset of parties' measuring devices are untrusted, it is accomplished by witnessing Einstein-Podolsky-Rosen (EPR) steering \cite{Schr,WJD07}, through the violation of a steering inequality \cite{ECavPRA09}. In both cases, if either the measurements performed would have been compatible, or the state would have been separable, then the corresponding violation could not have been obtained. The connection between measurement incompatibility and the violation of the Clauser-Horn-Shimony-Holt Bell inequality was first considered in \cite{Fine,Wolf}. Recently it was shown that there is a very strong relationship between a notion of incompatibility of measurements known as \emph{joint measurability} and EPR steering: a set of measurements are not jointly measurable if and only if they can be used to demonstrate steering \cite{steering vs JM,steering vs JM2}.

Besides their fundamental interest, scenarios involving a lack of trust are also of practical importance when the provider of the devices -- the source of entanglement or the measuring devices -- are untrustworthy. This is the situation which naturally arises in quantum cryptographic scenarios, where the provider could be an eavesdropper who naturally wants to break the cryptosystem. It is thus crucial to develop ways to certify steering or nonlocality in practical tests, where both the measurements and states are naturally noisy. In fact, loss-tolerant steering tests have been derived in the case of qubits using precise arrangement of measurement directions  \cite{lossy steering2,lossy steering3,lossy steering4,exp steering1}, and detection loophole free steering tests have already been performed \cite{exp steering1,exp steering2,exp steering3}.

Here we show how to construct loss-tolerant linear steering inequalities which are violated using \emph{any set} of $n$ von Neuman measurements as long as the detection efficiency of the test satisfies $\eta > 1/n$, where $\eta$ is the probability that the detector clicks. Crucially our construction works for all finite $n$, arbitrary dimension $d$ and for any choice of von Neumann measurements. Furthermore, the construction relies only upon the maximal overlap between any two measurement outcomes (of different measurements), a property which can easily be calculated for any finite set. Finally we show that by considering mutually unbiased basis (MUB) measurements in dimension $d$ (i) the violation of the inequalities can also tolerate arbitrary amounts of depolarising noise as $d$ increases (ii) can produce unbounded violations even in the presence of losses. Altogether, this should make the construction particularly relevant to experimental demonstrations of loophole free steering, especially those using higher-dimensional systems.

Finally, using the relation between EPR steering and joint measurability, our results further provide an explicit certificate that $n$ inefficient von Neumann measurements are not jointly measurable whenever $\eta > 1/n$. This matches the lower bound below which these measurements cannot demonstrate steering \cite{exp steering2}. 

The paper is organised as follows. We first introduce the relevant notions  of EPR steering and then show how to construct a steering inequality starting from any set of von Neuman measurements. We then show that this inequality witnesses steering for detection efficiencies satisfying $\eta > 1/n$ and study the tolerance to white noise when using MUB measurements. Finally we briefly introduce joint-measurability and show how the inequalities also witness the non-joint measurability of inefficient measurements.

\sectionprl{EPR steering and entanglement detection}
In an EPR steering test, we consider that Alice and Bob pre-share an unknown quantum state $\kets{\psi}{{\rmA\rmB}}$ onto which Alice performs one out of $n$ unknown measurements $\{M_{a|x}\}_x$ (labelled by $x = 1,\ldots, n$), each with $d$ outcomes (labelled by $a = 1,\ldots, d$). The unnormalised post-measurement states prepared for Bob are given by
\begin{equation}\label{e:assemblage}
\assem{\sigma}{a|x}{} = \tr_\rmA \left(\left(M_{a|x}\otimes \openone_\rmB\right)\proj{\psi}{{\rmA\rmB}}\right).
\end{equation}
The set $\{\assem{\sigma}{a|x}{\rmB}\}_{ax}$ is called an \textit{assemblage} \cite{Pusey}, from which one can obtain both the conditional probabilities $P(a|x)$ for Alice to obtain the outcome $a$ given that she made measurement $x$, $P(a|x) = \tr (\assem{\sigma}{a|x}{})$, as well as the conditional states themselves $\assem{\hat{\sigma}}{a|x}{} = \assem{\sigma}{a|x}{}/P(a|x)$. Bob is assumed to perform trusted measurements and therefore can perform full state tomography to determine to arbitrary accuracy the members of the assemblage he holds.

The interest in EPR steering derives from the fact that it allows one to certify the presence of quantum entanglement in this semi-device-independent scenario \cite{WJD07}.
To see how, let us assume that the source distributed a separable state $\rho^{\rmA\rmB} = \sum_i p_i \rho^\rmA_i\otimes\rho_i^\rmB$. This imposes the following structure on the assemblages created:
\begin{align}\label{e:LHS assem}
\assem{\sigma}{a|x}{} &= \sum_i p_i \tr_\rmA(M_{a|x}\rho_i^\rmA)\rho_i^\rmB,\nonumber\\
&= \sum_i q(a|x,i)\rho_i^\rmB,
\end{align}
where $q(a|x,i)=p_i \tr_\rmA(M_{a|x}\rho_i^\rmA)$. The above structure is called a \emph{Local Hidden State Model} (LHS model) for the assemblage $\{\assem{\sigma}{a|x}{}\}_{ax}$, and assemblages which have a LHS model are called unsteerable \cite{WJD07}.

Crucially, not all assemblages have a LHS model. This can always be certified through the violation of a linear steering inequality, given by
\begin{align}\label{e:steering inequality}
\beta = \sum_{ax}\tr\left(F_{a|x}{\sigma}_{a|x}\right) \leq \beta_\LHS,
\end{align}
where $\{F_{a|x}\}_{ax}$ is a collection of operators defining the inequality and
$\beta_\LHS$ is the maximum value that a unsteerable assemblage \eqref{e:LHS assem} can reach, \ie
 \begin{equation}\label{e:betaLHS}
 \beta_\LHS=\max_{\assem{\sigma}{a|x}{\LHS}  \in \LHS}\sum_{ax}\tr\left(F_{a|x}\assem{\sigma}{a|x}{\textsc{lhs}}\right).
 \end{equation}

\sectionprl{Lossy steering tests}
A crucial problem which arises when carrying out a steering test is the overall detection efficiency. That is, whereas in an idealised steering test the source will always create a pair of particles and in every run the measurements performed will give an outcome, in reality this is not the case. The particles may be lost en route, and the detectors may produce no click even if the a particle arrives. This problem becomes especially important in cryptographic applications, as an adversary can use the experimental imperfections to try and trick the parties into believing they have witnessed entanglement, although a separable state has in fact been used \cite{makarov1,makarov2}.

If the detection efficiency is not unity, instead of observing the assemblage \eqref{e:assemblage}, the one actually observed is given by
\begin{align}\label{e:assemblage eta}
\assem{\sigma}{a|x}{(\eta)} =
  \begin{cases}
   \eta \assem{\sigma}{a|x}{} &\text{for } a=1,\ldots,d \\
   (1-\eta)\assem{\sigma}{\rmR}{} &\text{for } a=\nc
  \end{cases}
\end{align}
where we have introduced the outcome $a = \nc$ to denote `no-click' events, $\assem{\sigma}{\rmR}{} = \sum_a \assem{\sigma}{a|x}{} = \tr_\rmA \proj{\psi}{{\rmA\rmB}}$ is the reduced state of Bob and $\eta = (1-P(\nc))$ is the overall detection efficiency of Alice, taking into account all of the losses, either on the channel or at the detectors. Following previous works \cite{lossy steering2,lossy steering3,lossy steering4,exp steering1,exp steering3,exp steering3} we do not consider the effects of losses in Bob´s side, 
since by assumption his devices are trusted and cannot be used by an eavesdropper to open the detection loophole.

Our goal is thus to derive steering inequalities which detect steering starting from assemblages of the form \eqref{e:assemblage eta}, whenever $\eta > 1/n$.

\sectionprl{Loss-tolerant steering inequalities}
Let us start by choosing an inequality formed of $n$ projective measurements of the form 
\begin{align}\label{e:inequality}
F_{a|x} =
  \begin{cases}
   \Pi_{a|x} &\text{for } a=1,\ldots,d \\
   \alpha\openone &\text{for } a=\nc
  \end{cases}
\end{align}
where each $\Pi_{a|x}$ is a rank-1 projector and at this stage $\alpha > 0$ is a positive constant which needs to be chosen in order to make the steering inequality \eqref{e:steering inequality} useful. It is necessary to determine $\beta_\LHS(\alpha)$, the LHS bound as a function of $\alpha$, which is found by maximising the value of $\beta$ over all LHS assemblages. This is seen to be the solution to the optimisation problem \eqref{e:betaLHS}.

As we show in the appendix, we can transform this problem in an instance of a semidefinite program \cite{boyd}, and exploit its duality theory to find a simple upper bound on $\beta_\LHS$. In particular, we show that by choosing 
\begin{align}\label{e:theta def}
\alpha &= \max_{x, x'>x, a, a'} \sqrt{\tr \left(\Pi_{a|x} \Pi_{a'|x'}\right)}\nonumber\\
&\equiv\cos\theta,
\end{align}
\ie the maximal inner product between any two measurement directions between any two different measurements, then
\begin{align}\label{e:beta LHS bound}
\beta_\LHS \leq 1+(n-1)\cos\theta.
\end{align}
Thus, any assemblage which obtains a value greater than this value demonstrates steering.

\sectionprl{Quantum violations}
We will now show that the above inequalities certify steering whenever $\eta > 1/n$. Assume that the assemblage in \eqref{e:assemblage eta} was created by Alice performing inneficient von Neumann measurements, \ie $M_{a|x} = \Pi_{a|x}$ on the maximally entangled state $\kets{\phi^+}{{\rmA\rmB}} = \sum_i \kets{i}{\rmA}\kets{i}{\rmB}/\sqrt{d}$. Consider furthermore the steering inequality of the form \eqref{e:inequality} with $F_{a|x} = \Pi_{a|x}^\intercal$ for $a\neq \nc$. A direct calculation shows that
\begin{equation}\label{e:beta quantum}
	\beta = \tr\sum_{ax}\assem{F}{a|x}{}\assem{\sigma}{a|x}{(\eta)} = n(\eta + (1-\eta)\cos\theta)
\end{equation}
where we use the fact that $\tr(A\otimes B \proj{\phi^+}{{\rmA\rmB}}) = \tr(AB^\intercal)$. The requirement $\beta > \beta_\LHS$ is satisfied whenever $\eta > 1/n$. We note that although in the above we considered that the maximally entangled state $\kets{\phi^+}{{\rmA\rmB}}$ is distributed between the parties, similar to \cite{steering vs JM} it is straightforward to adapt to a situation where an arbitrary pure Schmidt-rank-$d$ state $\kets{\psi}{{\rmA\rmB}} = \sum_i \sqrt{\lambda_i}\kets{i}{\rmA}\kets{i}{\rmB}$, with $\lambda_i > 0$ and $\sum_i \lambda_i = 1$, is distributed between the parties instead. Details can be found in the appendix.

In conclusion, any set of $n$ von Neumann measurements satisfying the minimal requirement that no two measurements share a common outcome can be used to demonstrate steering in a loophole-free manner as long as the detection efficiency satisfies $\eta > 1/n$, i.e. in a loss-tolerant manner.  A key advantage of our construction is that the specific choice of measurements is not important -- the only relevant property of the measurements used is the maximal overlap between any two distinct measurement outcomes, and as long as this is not unity \footnote{We note that this is not a restriction per se, since we can simply discard $m$ measurements, such that the remaining set of $(n-m)$ measurements contains no two measurements sharing a common direction.} then a steering inequality can be easily written down. 

\begin{figure}
\centering
\includegraphics[width=0.9\columnwidth]{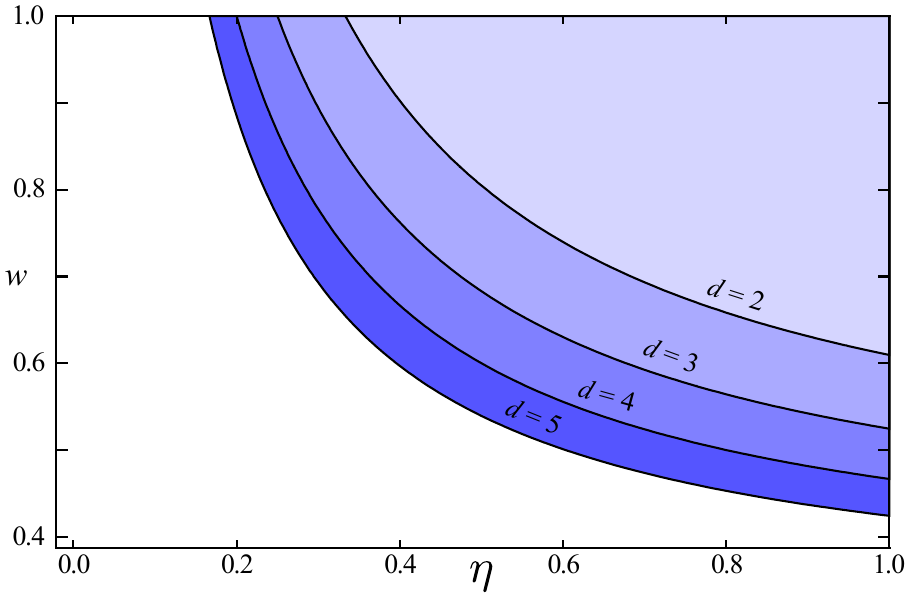}
\caption{(color online) Region plot of $w$ against $\eta$ for demonstrating steering using the inequality \eqref{e:inequality} with $d+1$ MUB measurements and the $d$ dimensional isotropic state \eqref{e:iso}. }\label{f:eta vs w}
\includegraphics[width=0.9\columnwidth]{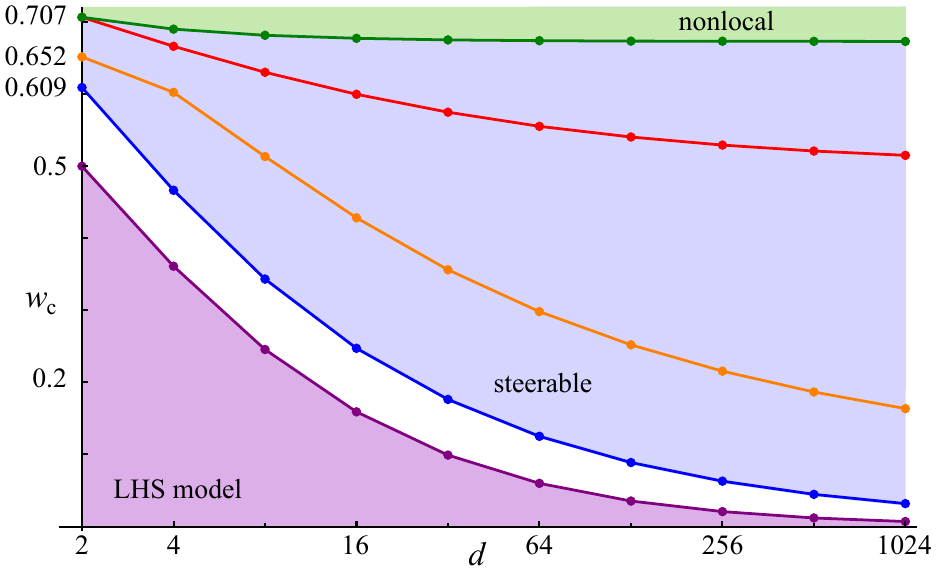}
\caption{(color online) Critical amount of white noise $w$ that can be tolerated by the isotropic state \eqref{e:iso} as a function of dimension for different steering and nonlocality tests. From bottom to top the curves refer to \textbf{Purple lowest curve:} bound below which a LHS model for projective measurements exists \cite{WJD07,Almeida}; \textbf{Blue curve:} critical $w$ derived using the steering inequality \eqref{e:inequality} with $d+1$ MUB measurements. Above this curve steering can be observed; \textbf{Orange curve:} bound obtained using the entropic steering inequality for $d+1$ MUB measurements derived in \cite{walborn}; \textbf{Red curve:} bound derived from the inequality \eqref{e:inequality} using only two MUB measurements; \textbf{Green highest curve:} bound obtained by the Collins-Gisin-Linden-Massar-Popescu Bell inequality \cite{CGLMP}, above which Bell nonlocality can be demonstrated. We have not plotted the bounds obtained from the inequalities derived in Ref. \cite{ECavPRA09} for spin measurements as they are only violated for $d=2$ and $3$.}\label{f:w vs d}
\end{figure}

\sectionprl{Robustness to white noise}
Since in practice one can never generate a pure maximally entangled state, it is important to see what level of noise can be tolerated by the steering tests presented here. To that end, let us consider that Alice and Bob share the isotropic state 
\begin{equation}\label{e:iso}
\rho(w) = w\proj{\phi^+}{{\rmA\rmB}} + (1-w)\openone/d^2
\end{equation} 
and that Alice performs projective measurements $\Pi_{a|x}$ as before. In this case the assemblage created is
\begin{align}
\assem{\sigma}{a|x}{(\eta,w)} =
  \begin{cases}
   \eta w \assem{\Pi}{a|x}{\intercal} + \frac{\eta}{d}(1-w)\frac{\openone}{d}&\text{for } a=1,\ldots,d \\
   (1-\eta)\frac{\openone}{d} &\text{for } a=\nc.
  \end{cases}
\end{align}
This leads to the following requirement to demonstrate steering
\begin{equation}
\eta > \frac{1}{n}\left(\frac{1-\cos \theta}{(1-\cos \theta) -(1-w)(1-1/d) }\right).
\end{equation}
Since $\cos \theta$ depends upon both $n$ and $d$ this bound is hard to analyse in general. However, let us specialise to the case of prime-power dimension and assume that Alice performs $n = d+1$ MUB measurements, in which case $\cos\theta = 1/\sqrt{d}$. For small $d$ we plot in Fig.~\ref{f:eta vs w} the region $\eta\times d$ for which steering can be demonstrated. This region grows in size with $d$, demonstrating the advantage of going beyond qubits. 

For large $d$ it is possible to calculate the asymptotic behaviour of the inequalities' violations. In particular, considering constant $w$,  the series expansion in $d$ leads to the following asymptotic behaviour for $\eta$:
\begin{equation}
\eta \geq \frac{1}{wd} + \mathrm{O}(d^{-3/2}).
\end{equation}
On the other hand, keeping $\eta$ constant, the asymptotic behaviour of $w$ for large $d$ is given by:
\begin{equation}
w \geq \frac{1}{\sqrt{d}} + \frac{1-\eta}{\eta d} + \mathrm{O}(d^{-3/2}).
\end{equation}

In Fig.~\ref{f:w vs d} we show the behaviour of the critical white noise tolerance $w_\mathrm{c}$ for $\eta = 1$, for exponentially growing system size $d$, and show the comparison with the the best known LHS bound and with other inequalities known for steering and nonlocality.

\sectionprl{Unbounded violations}
Another feature of our inequalities is that they allow to observe unbounded violations of steering inequalities which are also robust to losses. The study of unbounded violations of steering inequalities were recently initiated in  \cite{unbounded1,unbounded2} following on from the work which was done for the case of nonlocality \cite{unboundedNL1,unboundedNL2}, and was observed in a setting without losses. Following these works, we will define the \emph{normalised violation} of a steering inequality by $V = |\beta|/|\beta_\LHS|$, i.e. we are interested in the magnitude of the difference between the LHS bound and the quantum violation. 

From equations \eqref{e:beta LHS bound} and \eqref{e:beta quantum} we immediately see (recalling that by construction the steering inequality only takes non-negative values),
\begin{equation}
V \geq \frac{n(\eta + (1-\eta)\cos \theta)}{1+(n-1)\cos \theta}.
\end{equation} 
Specialising to the case of $d+1$ MUB measurements in prime-power dimension $d$, this becomes
\begin{equation}
V \geq \eta \sqrt{d} + O(1).
\end{equation}
Thus, whenever $\eta$ scales slower than $O(1/\sqrt{d})$, then an unbounded violation is obtained for sufficiently large dimension. This in particular includes the physically relevant case of constant losses independent of the dimension of the system. Note furthermore, that in the case $\eta =1$ we obtain exactly the same steering inequality as in \cite{unbounded2}, and therefore our construction can be seen as a generalisation of the one presented there, to include situations with losses. 
\sectionprl{Joint Measurability}
While for projective measurements the notion of compatibility of a collection of measurements is captured by the commutativity of the POVM elements, for more general measurements this is no longer adequate \cite{Busch}. The natural generalisation for general measurements is that of \emph{joint measurability}, which amounts to the existence of a single `parent POVM' from which, upon coarse-graining, all of the POVM elements can be obtained. More concretely, a set of $n$ $d$-outcome POVMs $\{\{M_{a|x}\}_a\}_x$ is said to be $n$-jointly measurable ($n$-JM) if there exists a single $n^d$ outcome parent POVM $\{M_\bfa\}_\bfa$, where $\bfa$ is an $n$-dit string $\rma_1\cdots \rma_n$ such that
\begin{align}\label{e:JM}
M_{\rma_x|x} &= \sum_{\bfa / \rma_x} M_{\bfa} & & \forall \rma_x,x
\end{align}
where $\bfa / \rma_x = \rma_1 \cdots \rma_{x-1} \rma_{x+1}\cdots \rma_n$ is the string formed of all the dits of $\bfa$ except $\rma_x$.

As mentioned in the introduction, EPR steering not only certifies the presence of entanglement, but also the presence of non-JM measurements, as it can straightforwardly be shown that measurements of the form \eqref{e:JM} when applied by Alice to half of any state prepare LHS assemblages for Bob, i.e. of the form \eqref{e:LHS assem}.

Consider now the set of inefficient von Neumann measurements $\assem{M}{a|x}{\eta}$ given by
\begin{align}\label{e:noisy measurements}
\assem{M}{a|x}{(\eta)} =
  \begin{cases}
   \eta \Pi_{a|x} &\text{for } a=1,\ldots,d \\
   (1-\eta)\openone &\text{for } a=\nc
  \end{cases}
\end{align}
where $P(\nc) = (1-\eta)$ is the probability of obtaining a `no-click' outcome labelled by $\nc$, and $\{\Pi_{a|x}\}_{x}$ is a von Neumann projective measurement for each $x$. These measurements can be seen to exactly prepare assemblages of the form \eqref{e:assemblage eta}. That is, we can think of these as the measurements actually being performed by Alice in a lossy steering test. Since we have seen in the previous section that \eqref{e:assemblage eta} demonstrates steering whenever $\eta > 1/n$, this certifies that the measurements \eqref{e:noisy measurements} are not JM for the same range of $\eta$. In the appendix we show a explicit  parent POVM for $\eta \leq 1/n$.

\sectionprl{Discussion}
We have given a general construction of experimentally friendly loss-tolerant linear steering inequalities in arbitrary dimension which are violated whenever Alice performs von Neumann measurements and the losses are not worse than $\eta > 1/n$. Moreover, the violation of these inequalities tolerates high values of white noise. Put together, these facts promote the inequalities derived here as strong experimental tests of EPR steering, that are robust to losses and experimental imperfections and are valid for any set of von Neumann measurements in any dimension. A fundamental consequence of our results is the fact that $n$ inneficient von Neumann measurements become jointly measurable only when their detection efficiencies are below $1/n$.

We would like to finish by comparing the results obtained here with the previous state-of-the art results concerning steering and Bell nonlocality tests with losses. 
In \cite{lossy steering2,lossy steering4,exp steering1} a steering test involving a two-qubit maximally entangled state and 16 measurements arranged in a precise way was shown to tolerate efficiencies down to $\eta>1/16$. The steering tests provided here go beyond this result in several senses: First, we have shown that any set of $n$ measurements suffices to demonstrate steering iff detection efficiencies are higher than $1/n$. Moreover, our construction works for states in every dimension. In fact, by increasing the dimension the present steering tests become arbitrarily robust to white noise (see Fig. \ref{f:w vs d}). Notice that Ref. \cite{lossy steering3} shows a steering test that tolerates arbitrary losses, although the state that has to be used in this test approaches a separable state, thus quite fragile to experimental imperfections. When it comes to Bell tests, the best known Bell inequalities \cite{VPB} need to increase the state's dimension to tolerate arbitrary losses, whereas for the steering tests developed here arbitrary losses can be tolerated in any dimension. Moreover, the inequalities of \cite{VPB} work, once more, for specific choices of states and measurements, and can only tolerate low levels of noise, again in contrast to those demonstrated here.

We thank A. Ac\'in, P. Kwiat, T. Graham and A. M\'attar for discussions. DC is supported by the Beatriu de Pin\'os fellowship (BP-DGR 2013) and PS by the Marie Curie COFUND action through the ICFOnest program and the ERC CoG QITBOX.

\appendix
\begin{appendix}
\section{Dual problem of $\beta_\textsc{LHS}$}Here we show that the problem (7) of the main text can be transformed into an SDP, and use the duality theory of semidefinite programming to derive its dual, which will allow us to find an upper bound on $\beta_\LHS$. Let us start by noting that we can re-write any unsteerable assemblage (Eq.~(2) of the main text) as 
\begin{align}
\assem{\sigma}{a|x}{} &= \sum_\bfa D_\bfa(a|x)\assem{\sigma}{\bfa}{}&& \forall a,x,
\end{align}
where $\bfa = \rma_1\cdots \rma_n$ is an $n$-dit string, $D_\bfa(a|x)= \delta_{a,\rma_x}$ are the deterministic single-party behaviours, whereby Alice outputs deterministically $a = \rma_x$ when her input is $x$, and $\assem{\sigma}{\bfa}{} = \sum_i p_i p(\bfa|i)\rho_i^\rmB$. This form is advantageous, since the $D_\bfa(a|x)$ are fixed, thus this takes the form of a set of linear matrix inequalities (LMIs). Using this re-writing, the problem (7) of the main text can be written as
\begin{align}
\beta_\LHS= \max_{\sigma_{\bfa}}&\quad \tr \sum_{ax\bfa}\assem{F}{a|x}{}D_\bfa(a|x)\assem{\sigma}{\bfa}{} \nonumber \\
\text{s.t.}&\quad \tr\sum_\bfa \assem{\sigma}{\bfa}{} = 1, \\
&\quad \assem{\sigma}{\bfa}{} \geq 0\quad \forall \bfa. \nonumber
\end{align}
which is now seen to be an SDP, since all constraints are positive semidefinite constraints or LMIs. To obtain the dual, we first write the Lagrangian of this problem
\begin{eqnarray}
\mathcal{L}&=&\tr \sum_{ax\bfa}\assem{F}{a|x}{}D_\bfa(a|x)\assem{\sigma}{\bfa}{}+\sum_{\bfa}H_{\bfa}\assem{\sigma}{\bfa}{}+\gamma(1-\tr\sum_{\bfa}\assem{\sigma}{\bfa}{})\nonumber\\
&=&\tr \sum_{\bfa}\assem{\sigma}{\bfa}{}\left[\sum_{ax}\assem{F}{a|x}{}D_{\bfa}(a|x)+H_{\bfa}-\gamma\identity\right]+\gamma,
\end{eqnarray}
where $\gamma$ and $\{H_{\bfa}\}_\bfa$ are the dual variables to the first and second (sets of) constraints respectively. This Lagrangian is unbounded from above unless
\begin{equation}
\sum_{ax}\assem{F}{a|x}{}D_{\bfa}(a|x)+H_{\bfa}-\gamma\identity=0\quad\forall \bfa.
\end{equation}
Imposing this constraint enforces $\mathcal{L}=\gamma$. By choosing $H_{\bfa}\geq0$ we have that the following minimization problem upper bounds the primal problem $\beta_\LHS$:
\begin{align}
\gamma^* = \min&\quad \gamma \nonumber \\
\text{s.t.}&\sum_{ax}\assem{F}{a|x}{}D_{\bfa}(a|x)+H_{\bfa}-\gamma\identity=0\quad\forall \bfa\nonumber\\
&H_{\bfa}\geq0,
\end{align}
which can be simplified to
\begin{align}\label{e:dual app}
\gamma^* = \min&\quad \gamma \nonumber \\
\text{s.t.}&\quad \sum_{ax}\assem{F}{a|x}{}D_\bfa(a|x) \leq \gamma \identity \quad \forall \bfa.
\end{align}

\sectionprl{Proof of upper bound on $\gamma^*$}
 By defining $G_\bfa = \sum_{ax}F_{a|x}D_\bfa(a|x) \equiv \sum_{x}F_{\rma_x|x}$, problem \eqref{e:dual app} is seen to be equal to
\begin{equation}
\gamma^* = \max_\bfa \|G_\bfa\|_\infty
\end{equation}
where the maximisation is over the set of $(d+1)^n$ operators $G_\bfa$, one corresponding to each deterministic strategy which Alice can employ. Let us define for each $\bfa$ the number of no-click outcomes $\nc$ that the string contains, which we denote by  $|\bfa|_\nc$. Given this, we can split the set $\{G_\bfa\}_\bfa$ into sets $H_k$ according to the number of no-click outcomes,
\begin{equation}
H_k = \{G_\bfa \big| |\bfa|_\nc = k\}
\end{equation}
The purpose for doing this is that now each $G_\bfa$ inside the set $H_k$ has the same structure, namely
\begin{equation}
G_\bfa = \sum_{i=1}^{n-k} \Pi_i + k\alpha \openone
\end{equation}
where we denote an arbitrary rank-1 projector as $\Pi_i$. Therefore, inside each set $H_k$ the operator norm of each member is given by
\begin{equation}
\|G_\bfa\|_\infty =\|\Pi_1 + \cdots + \Pi_{n-k}\|_\infty + k\alpha
\end{equation}
To proceed we make use of the following result, which will be proved in the proceeding section:
\begin{align}
\|\Pi_1 + \cdots + \Pi_\ell\|_\infty &\leq 1 + (\ell - 1)\cos\varphi\nonumber \\
\cos \varphi &= \max_{i, j>i} \|\Pi_i \Pi_j\|_\infty
\end{align}
Since each $G_\bfa = \sum_x F_{\rma_x|x}$ contains only at most one measurement direction from each measurement $\Pi_{a|x}$ it is clear that if we define
\begin{align}
\cos\theta &= \max_{x, x'>x, a, a'} \| \Pi_{a|x} \Pi_{a'|x'}\|_\infty \nonumber \\
&= \max_{x, x'>x, a, a'} \sqrt{\tr \left(\Pi_{a|x} \Pi_{a'|x'}\right)}
\end{align}
then $\cos \theta \geq \cos\varphi$, since $\cos\varphi$ comes from taking a maximisation over a subset of the set maximised over for $\cos \theta$. Thus, we obtain an upper bound for every $\|G_\bfa\|_\infty$ depending only upon the set $H_k$ it belongs to
\begin{align}
\|G_\bfa\|_\infty \leq k\alpha + (n-k-1)\cos\theta, &&\forall G_\bfa \in H_k
\end{align}
valid except when $k = n$. In this exceptional case, consisting of the single strategy $\bfa = \nc\cdots\nc$ of $n$ no-click outcomes, $G_{\nc\cdots\nc} = n \alpha \openone$ and $\|G_{\nc\cdots\nc}\|_\infty = n \alpha $ by inspection.

We thus finally see that by choosing $\alpha = \cos\theta$ we have
\begin{align}
\|G_\bfa\|_\infty \leq
  \begin{cases}
   1 + (n-1)\cos \theta &G_\bfa \notin H_n \\
   n\cos \theta &G_\bfa \in H_n
  \end{cases}
\end{align}
Thus whenever $\cos\theta < 1$, i.e. when two measurements do not share an outcome then $n > 1 + (n-1)\cos \theta > n\cos\theta$. We thus have
\begin{equation}
\beta_\LHS \leq 1 + (n-1)\cos \theta
\end{equation}
as the desired result. We end by noting that $(\alpha = 1, \beta_\LHS = n)$ corresponds to a trivial inequality which can never be violated, which is seen for example by considering the assemblage created from perfectly efficient measurements ($\eta = 1$) on the maximally entangled state, which obtains the value $n = \beta_\LHS$.

\sectionprl{Bounding the norm of $k$ rank-1 projectors}
In this section we will prove the following inequality which holds for the sum of $\ell$ rank-1 projectors acting on an arbitrary finite dimensional Hilbert space $\mathbb{C}^d$.
\begin{align}
\|\Pi_1 + \cdots + \Pi_\ell\|_\infty &\leq 1 + (\ell - 1)\cos\varphi\nonumber \\
\cos \varphi &= \max_{i, j>i} \|\Pi_i \Pi_j\|_\infty
\end{align}
Let us introduce an auxiliary Hilbert space $\mathbb{C}^\ell$, and define a standard basis $\ket{i}$, $i = 1, \ldots \ell$, for this space. Consider then the operator $X$ acting on $\mathbb{C}^\ell \otimes \mathbb{C}^d$
\begin{equation}
X = \sum_i \ket{1}\bra{i}\otimes \Pi_i
\end{equation}
which is nothing but a block matrix, with the first block-row containing the projectors $\Pi_i$. First we will use the fact that $\|X^\dagger X\|_\infty = \|XX^\dagger\|_\infty$. We see that
\begin{align}
XX^\dagger &=\ket{1}\bra{1}\otimes \sum_i \Pi_i,\nonumber\\ X^\dagger X &= \sum_{ij}\ket{i}\bra{j}\otimes \Pi_i \Pi_j
\end{align}
Clearly $\|XX^\dagger\|_\infty = \|\Pi_1 + \cdots + \Pi_\ell\|_\infty$ is what we desire to bound. Therefore we will use $X^\dagger X$ to do so. First, we note that we can write
\begin{equation}\label{e:decom}
X^\dagger X = \sum_{i}\ket{i}\bra{i}\otimes \Pi_i
+ \sum_{j=1}^{\ell-1}\sum_i \ket{i}\bra{i\oplus j}\otimes \Pi_i\Pi_{i\oplus j}
\end{equation}
where $\oplus$ denotes addition modulo $\ell$. This decomposition amounts to writing $X^\dagger X$ as a block diagonal matrix plus a sum of $\ell -1$ matrices, each with a block structure and containing only displaced diagonals (i.e. have the structure of a block permutation matrix).

The first term of the right hand side of \eqref{e:decom} has operator norm
\begin{equation}
\left\|\sum_{i}\ket{i}\bra{i}\otimes \Pi_i\right\|_\infty = \max_i \|\Pi_i\|_\infty = 1
\end{equation}
since the operator norm of a block diagonal operator is the maximal operator norm of any block, which in our case is unity. For each of the remaining terms we can use the fact that the operator norm, being equal to the largest singular value, is invariant under the transformation $X\to UXV$ where $U$ and $V$ are unitary. Choosing $U = \openone \otimes \openone$ and $V_j = \sum_i \ket{i\oplus j}\bra{i} \otimes \openone$ we see that
\begin{align}
U\sum_i \ket{i}\bra{i\oplus j}\otimes \Pi_i\Pi_{i\oplus j}V_j = \sum_i \ket{i}\bra{i}\otimes \Pi_i\Pi_{i\oplus j}
\end{align}
and thus
\begin{align}
\left\|\sum_{i}\ket{i}\bra{i\oplus j}\otimes \Pi_i\Pi_{i\oplus j}\right\|_\infty = \max_i \|\Pi_i\Pi_{i\oplus j}\|_\infty
\end{align}
again due to the block structre of the transformed matrix. Since $\max_i \|\Pi_i\Pi_{i\oplus j}\|_\infty \leq \max_{i,j>i} \|\Pi_i\Pi_{j}\|_\infty = \cos \varphi$ we can place the same bound $\cos\varphi$ on each of the $\ell - 1$ terms. Finally by using repeatedly the triangle inequality $\|X+Y\|_\infty \leq \|X\|_\infty + \|Y\|_\infty$ be obtain the desired result
\begin{align}
\|\Pi_1 + \cdots + \Pi_\ell\|_\infty &\leq 1 + (\ell - 1)\cos\varphi\nonumber
\end{align}
\sectionprl{Quantum violations with arbitrary pure entangled states} In this section we will show that it is possible to use an arbitrary pure entangled  Schmidt-rank $d$ state to demonstrate steering with arbitrary losses. In particular let us assume that the state $\kets{\psi}{{\rmA\rmB}} = \sum_i \sqrt{\lambda_i}\kets{i}{\rmA}\kets{i}{\rmB}$, with $\lambda_i > 0$ and $\sum_i \lambda_i = 1$, is distributed between the parties. Defining the matrix $D = \sum_i \sqrt{d\lambda_i}\ket{i}\bra{i}$, then we have that $\kets{\psi}{{\rmA\rmB}} = D\otimes \openone\kets{\phi^+}{{\rmA\rmB}}$, i.e. we can see it as the (unnormalised state) after a local filtering by Alice. By performing measurements on this state, Alice prepares the assemblage
\begin{align}
\assem{\sigma}{a|x}{(\eta)} =
  \begin{cases}
   \eta (D\assem{\Pi}{a|x}{}D)^\intercal &\text{for } a=1,\ldots,d \\
   (1-\eta)\assem{\sigma}{\rmR}{} &\text{for } a=\nc
  \end{cases}
\end{align}
where $\sigma_\rmR = D^2/d$ is the reduced state of Bob. Notice that $(D\assem{\Pi}{a|x}{}D)^\intercal = p'(a|x) \assem{\Pi}{a|x}{'}$, i.e. the action of $D$ does not stop Alice from preparing rank-1 states for Bob, only the directions and normalisations have changed. It thus follows directly that the inequality
\begin{align}
F'_{a|x} =
  \begin{cases}
   \Pi'_{a|x} &\text{for } a=1,\ldots,d \\
   \cos\theta'\openone &\text{for } a=\nc
  \end{cases}
\end{align}
where 
\begin{align}
\cos\theta' &= \max_{x, x'>x, a, a'} \sqrt{\tr \left(\Pi'_{a|x} \Pi'_{a'|x'}\right)}
\end{align}
certifies steering as long as $\cos\theta' < 1$ and $\eta > 1/n$.

\sectionprl{Joint measurability of $n$ inefficient measurements when $\eta \leq 1/n$}
In this section we show that any set of $n$ inefficient measurements (von Neumann or not, with an arbitrary number of outcomes) is jointly measurable if $\eta \leq 1/n$. Consider the $n$ measurements $\{M_{a|x}\}_x$ for $x = 1,\ldots, n$ and $a = 1,\ldots, d$. The inefficient measurements formed from this set are $\{M^{(1/n)}_{a|x}\}_x$
\begin{align}\label{e: 1/n meas}
\assem{M}{a|x}{(1/n)} =
  \begin{cases}
   \frac{1}{n} M_{a|x} &\text{for } a=1,\ldots,d \\
   (1-\frac{1}{n})\openone &\text{for } a=\nc
  \end{cases}
\end{align}
Consider now the parent POVM $\{M_\bfa\}_\bfa$, with $\bfa$ an $n$ $(d+1)$-valued string given by
 \begin{align}
M_\bfa =
  \begin{cases}
   \frac{1}{n} M_{\rma_x|x} &\text{if } |\bfa|_\nc = (n-1) \text{ and } \rma_x \neq \nc \\
  0 &\text{if } |\bfa|_\nc\neq(n-1)
  \end{cases}
\end{align}
That is, all but $n(d+1)$ of the $n^{d+1}$ POVM elements vanish, the remaining corresponding to giving `no-click' outcomes to $(n-1)$ measurements and giving an actual outcome for the remaining one. It is clear that evaluting (16) of the main text for this parent POVM that we recover the inefficient measurements \eqref{e: 1/n meas}.
\end{appendix}
\end{document}